\documentclass[aps,prl,twocolumn,superscriptaddress,showpacs]{revtex4}
\usepackage{amsfonts}
\usepackage{epsfig}
\usepackage{amsmath,amssymb}
\usepackage{times}

\begin{document}

\title{Noise bridges dynamical correlation and
topology in coupled oscillator networks}

\author{Jie Ren}
\affiliation{ NUS Graduate School for Integrative Sciences and
Engineering, Singapore 117456, Republic of Singapore }
\affiliation{ Department of Physics and Centre for Computational
Science and Engineering, National University of Singapore,\\
Singapore 117546, Republic of Singapore }

\author{Wen-Xu Wang}
\affiliation{School of Electrical, Computer, and Energy
Engineering, Arizona State University, Tempe, AZ 85287, USA}

\author{Baowen Li}
\affiliation{ NUS Graduate School for Integrative Sciences and
Engineering, Singapore 117456, Republic of Singapore }
\affiliation{ Department of Physics and Centre for Computational
Science and Engineering, National University of Singapore,\\
Singapore 117546, Republic of Singapore }

\author{Ying-Cheng Lai}
\affiliation{School of Electrical, Computer, and Energy
Engineering, Arizona State University, Tempe, AZ 85287, USA}
\affiliation{Department of Physics, Arizona State University,
Tempe, AZ 85287, USA}

\date{\today}

\begin{abstract}
We study the relationship between dynamical properties and
interaction patterns in complex oscillator networks in the
presence of noise. A striking finding is that noise leads to a
general, one-to-one correspondence between the dynamical
correlation and the connections among oscillators for a variety of
node dynamics and network structures. The universal finding
enables an accurate prediction of the {\em full network topology}
based solely on measuring the dynamical correlation. The power of
the method for network inference is demonstrated by the high
success rate in identifying links for distinct dynamics on both
model and real-life networks. The method can have potential
applications in various fields due to its generality, high
accuracy and efficiency.
\end{abstract}
\maketitle

Understanding the relationship between dynamics and network
structure is a central issue in interdisciplinary science
\cite{Review3,Review4}. Despite the tremendous efforts in
revealing the topological effect on a variety of dynamics
\cite{CN_spreading,CN_syn,CN_vib}, how to infer the interaction
pattern from dynamical behaviors is still challenging as an
inverse problem, especially in the absence of the knowledge of
nodal dynamics. Some methods aiming to address the inverse problem
have been proposed, such as spike classification methods for
measuring interactions among neurons from spike trains \cite{NN},
and approaches based on response dynamics \cite{Timme:2007}, $L1$
norm \cite{NS:2008} and noise scaling \cite{WCHLH:2009}. For the
inverse problem, a basic question is whether sufficient
topological information can be obtained from measured time series
of dynamics. In this regard, the answer is negative when there is
strong synchronization as, in this case, the coupled units behave
as a single oscillator and interactions among units vanish so that
it is impossible to extract the interaction pattern from
measurements.

Quite surprisingly, we find that with the help of noise, in
general it becomes possible to precisely identify interactions
based solely on the correlations among measured time series of
nodes. In this sense, we say that {\em noise bridges dynamics and
topology}, facilitating inference of network structures. We note
that noise is ubiquitous in physical and natural systems and
understanding the noise effect on dynamical systems has been a
fundamental issue in nonlinear and statistical physics. While
there are recent works on the interplay between collective
dynamics and topology of complex systems under noise
\cite{MMG:2007,TF:2008,MGLM:2008} and on predicting node degrees
for complex networks \cite{WCHLH:2009}, taking advantage of noise
to predict the {\em full connecting topology} of an unknown
complex network is an outstanding question. Addressing this
question not only is fundamental to nonlinear science, but also
can have significant applications in diverse areas such as
computer networks, biomedical systems, neuroscience,
socio-economics, and defense.

In this Letter, we present a general and powerful method to
precisely identify links among nodes based on the noise-induced
relationship between dynamical correlation and topology. 
 Analytically, we find that there exists
a one-to-one correspondence between the dynamical correlation
matrix of nodal time series and the connection matrix of
structures, due to the presence of noise. This finding enables an
accurate prediction of network topology from time series.
Numerical simulations are performed using four typical dynamical
systems, together with several model and real networks. For all
cases examined, comparisons between the original and the predicted
topology yield uniformly high success rate of prediction. The
advantages of our noise-based method are then: (i) high accuracy
and efficiency, (ii) generality with respect to node dynamics and
network structures, (iii) no need for control, and (iv)
applicability even when there is weak coherence in the collective
dynamics.

Our general approach to bridging dynamical correlation and
topology is, as follows. We consider $N$ {\em nonidentical}
oscillators, each of which satisfies $\dot{\mathbf{x}}_i =
\mathbf{F}_i(\mathbf{x}_i)$ in the absence of coupling, where
$\mathbf{x}_i$ denotes the $d$-dimensional state variable of the
$i$th oscillator. Under noise, the dynamics of the whole
coupled-oscillator system can be expressed as:
\begin{equation} \label{eq:master}
\dot{\mathbf{x}}_i = \mathbf{F}_i(\mathbf{x}_i) - c \sum^N_{j=1}
L_{ij} \mathbf{H}(\mathbf{x}_j) + \eta_i,
\end{equation}
where $c$ is the coupling strength, $\mathbf{H}: \mathbb{R}^d
\rightarrow \mathbb{R}^d$ denotes the coupling function of
oscillators, $\eta_i$ is the noise term, $L_{ij} = -1$ if $j$
connects to $i$ (otherwise $0$) for $i \neq j$ and $L_{ii} =
-\sum_{j=1,j\neq i}^{N}L_{ij}$. Due to nonidentical oscillators
and noise, an invariant synchronization manifold does not exist.
Let $\mathbf{\bar{x}}_i$ be the counterpart of $\mathbf{x}_i$ in
the absence of noise, and assume a small perturbation $\xi_i$, we
can write $\mathbf{x}_i = \mathbf{\bar{x}}_i + \xi_i$.
Substituting this into Eq. (\ref{eq:master}), we obtain:
\begin{equation} \label{eq:linearize}
\dot{\xi} = [D\mathbf{\hat{F}}(\mathbf{\bar{x}}) - c
\mathbf{\hat{L}} \otimes D\mathbf{\hat{H}}(\mathbf{\bar{x}}) ]\xi
+ \eta,
\end{equation}
where $\xi = [\xi_1,\xi_2, \ldots , \xi_N ]^T$ denotes the
deviation vector, $\eta = [\eta_1,\eta_2,\ldots, \eta_N]^T$ is the
noise vector, $\mathbf{\hat{L}}$ names the Laplacian matrix of
coupling $\{L_{ij}\}$, $D\mathbf{\hat{F}}(\mathbf{\bar{x}}) =
\mathrm{diag}
[D\mathbf{\hat{F}}_1(\mathbf{\bar{x}}_1),D\mathbf{\hat{F}}_2(\mathbf{\bar{x}}_2),\cdots
, D\mathbf{\hat{F}}_N(\mathbf{\bar{x}}_N)]$ ($D\mathbf{\hat{F}}_i$
are $d\times d$ Jacobian matrices of $\mathbf{F}_i$), $\otimes$
denotes direct product, and $D\mathbf{\hat{H}}$ is the Jacobian
matrix of the coupling function $\mathbf{H}$.

Denoting the dynamical correlation of oscillators $\langle \xi
\xi^T \rangle$ as $\mathbf{\hat{C}}$, wherein $C_{ij} = \langle
\xi_i \xi_j \rangle$ and $\langle \cdot \rangle$ is time average,
we have
\begin{equation} \label{eq:correlation}
0 =\langle d(\xi \xi^T)/dt\rangle = -\mathbf{\hat{A}\hat{C}} -
\mathbf{\hat{C}}\mathbf{\hat{A}}^T + \langle \eta \xi^T \rangle +
\langle \xi \eta^T \rangle,
\end{equation}
where $\mathbf{\hat{A}} = -D\mathbf{\hat{F}}(\mathbf{\bar{x}}) + c
\mathbf{\hat{L}} \otimes D\mathbf{\hat{H}}(\mathbf{\bar{x}})$. To
obtain the expression of $\langle \eta \xi^T \rangle$ and $\langle
\xi \eta^T \rangle$, we get the solution $\xi(t)$ from
Eq.~(\ref{eq:linearize}): $\xi(t)= \mathbf{\hat{G}}(t-t_0)\xi(t_0)
+ \int_{t_0}^{t}dt'\mathbf{\hat{G}}(t-t')\eta(t'), $ where
$\mathbf{\hat{G}}(t) = \exp(-\mathbf{\hat{A}}t)$. In the absence
of divergence of state variables, $\mathbf{\hat{G}}(\infty) = 0$.
Setting $t_0 \rightarrow -\infty$, without loss of generality, we
have $\xi(t)= \int_{-\infty}^{t}\mathbf{\hat{G}}(t-t')\eta(t')
dt'$. Note that $\mathbf{\hat{G}}(0) = \mathbf{\hat{I}}$, we hence
obtain $\langle \xi \eta^T \rangle =
\int_{-\infty}^{t}\mathbf{\hat{G}}(t-t')\langle
\eta(t)\eta^T(t')\rangle dt' \nonumber  =
\int_{-\infty}^{t}\mathbf{\hat{G}}(t-t')\mathbf{\hat{D}}\delta(t-t')
dt' = \mathbf{\hat{D}}/2, $ where $\mathbf{\hat{D}}$ is the
covariance matrix of noise. Analogously, we can obtain $\langle
\eta \xi^T \rangle = \mathbf{\hat{D}}/2$. Therefore,
Eq.~(\ref{eq:correlation}) can be simplified to:
\begin{equation} \label{eq:general}
\mathbf{\hat{A}\hat{C}}+\mathbf{\hat{C}}\mathbf{\hat{A}}^T =
\mathbf{\hat{D}}.
\end{equation}
Since $\mathbf{\hat{A}} = -D\mathbf{\hat{F}}(\mathbf{\bar{x}}) + c
\mathbf{\hat{L}} \otimes D\mathbf{\hat{H}}(\mathbf{\bar{x}})$, the
above equality reveals a general relationship between the
dynamical correlation $\mathbf{\hat{C}}$ and the connecting matrix
$\mathbf{\hat{L}}$ in the presence of noise as characterized by
$\mathbf{\hat{D}}$. The general solution of $\mathbf{\hat{C}}$ can
be written as $\mbox{vec}(\mathbf{\hat{C}})
=\mbox{vec}(\mathbf{\hat{D}})/(\mathbf{\hat{I}}\otimes
\mathbf{\hat{A}} + \mathbf{\hat{A}} \otimes \mathbf{\hat{I}})$,
where $\mbox{vec}(\mathbf{\hat{X}})$ is a vector containing all
columns of matrix $\mathbf{\hat{X}}$ \cite{matrix}.

\begin{figure}
\begin{center}
\epsfig{figure=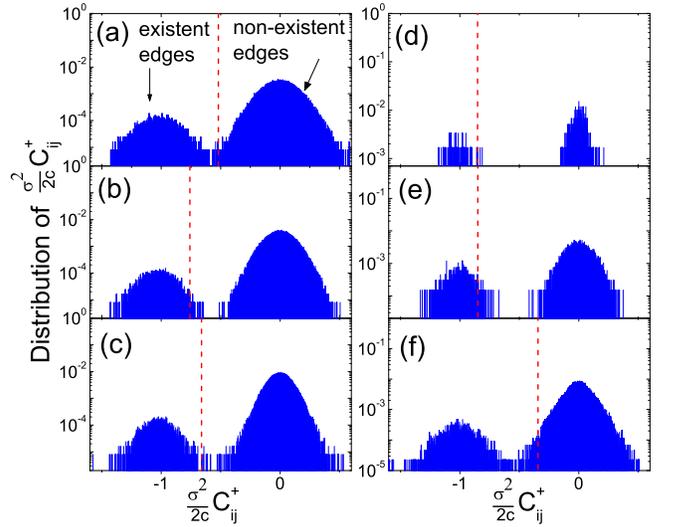,width=\linewidth} \vspace{-.5cm}
\caption{(Color online) Distribution of the values of
$[\sigma^2/(2c)]C_{ij}^\dag$, where $C_{ij}^\dag$ are the elements
in the pseudo inverse matrix of the dynamical correlation matrix
$\mathbf{\hat{C}}$. Consensus dynamics \cite{Saber:2007} are used
for (a) random \cite{ER:1959}, (b) small-world \cite{WS:1998}, (c)
scale-free model networks \cite{BA:1999} and three real-world
networks: (d) friendship network of karate club \cite{FKC}, (e)
network of American football games among colleges \cite{ACF} and
(f) the neural network of C. Elegans \cite{WS:1998}. The
theoretical threshold $[\sigma^2/(2c)]C_{M}^\dag$ is marked by red
dashed lines. The sizes of model networks are all 500. For random
networks, the connection probability among nodes is $0.024$. For
scale-free networks the minimum degree is $k_{min}=6$. For
small-world networks, $\langle k\rangle =12$ and the rewiring
probability is 0.1.} \label{fig:distribution}
\end{center}
\end{figure}

For illustrative purpose, we consider one-dimensional state
variable and linear coupling such that $D\mathbf{\hat{H}}=1$, with
Gaussian white noise $\mathbf{\hat{D}}=\sigma^2 \mathbf{\hat{I}}$,
and further regard the intrinsic dynamics $D\mathbf{\hat{F}}$ as
small perturbations. Then Eq.~(\ref{eq:general}) can be simplified
to $\mathbf{\hat{L}\hat{C}} + \mathbf{\hat{C}}\mathbf{\hat{L}}^T =
\sigma^2\mathbf{\hat{I}}/c$. For an undirected network with
symmetric coupling matrix, the solution of $\mathbf{\hat{C}}$ can
be expressed as:
\begin{equation} \label{eq:undirect}
\mathbf{\hat{C}} = \frac{\sigma^2}{2c}\mathbf{\hat{L}}^\dag,
\end{equation}
where $\mathbf{\hat{L}}^\dag$ denotes the pseudo inverse of the
Laplacian matrix. We note that the dynamic correlation matrix
$\mathbf{\hat{C}}$ is closely related to the network connection
matrix $\mathbf{\hat{L}}$, which can be used to infer network
structures when no knowledge about the nodal dynamics is
available. In fact, $\mathbf{\hat{C}}$ acts as the ``Green's
function'' of the network and can be expressed as some kind of
path integral associated with the underlying network topology (see
\cite{note:green}), as follows:
\begin{equation}
C_{ij} = \frac{\sigma^2}{2c}\sum_{\mathbf{path}} \prod_{m \in
\mathbf{path} }\frac{1}{k_m},
\end{equation}
where $\mathbf{path}$ means all paths from $j$ to $i$, and $m$
denotes the nodes on them. This path-integral representation is
extremely useful for revealing the direct relation between
autocorrelation $C_{ii}$ in the matrix $\mathbf{\hat{C}}$ and the
local structure $k_i$. In particular, for $n$th-order
approximation, we count all paths whose lengths are equal to or
less than $n$. Under second-order approximation, we have
\begin{equation}
C_{ii} = \frac{\sigma^2}{2c}\left( \frac{1}{k_i} + \frac{1}{k_i^2}
\sum_{q \in \Gamma_i} \frac{1}{k_q} \right) \simeq
\frac{\sigma^2}{2ck_i}\left(1+ \frac{1}{\langle k\rangle}\right),
\label{eq:local}
\end{equation}
where mean-field approximation is applied and $\Gamma_i$ denote
the neighbors of node $i$. This dependence of the autocorrelation
$C_{ii}$ on the degree $k_i$, under the second-order approximation
is consistent with the recently discovered noise-induced algebraic
scaling law in Ref. \cite{WCHLH:2009}, derived there by a
power-spectral analysis.

To provide numerical support for the validity and generality of
our theoretical results on the relationship between dynamical
correlation and topology, we consider a number of model and
real-world network structures by using four typical dynamical
systems, as follows. (i) \emph{Consensus dynamics}
\cite{Saber:2007}: $\dot{x}_i = c\sum_{j=1}^{N}P_{ij}(x_j - x_i)
+\eta_i$; (ii) \emph{Identical R\"ossler dynamics}
\cite{Rossler:1976} (I-R\"ossler): $\dot{x}_i=-y_i-z_i +
c\sum_{j=1}^{N}P_{ij}(x_j- x_i) + \eta_i$, $\dot{y}_i = x_i +
0.2y_i + c\sum_{j=1}^{N} P_{ij}(y_j - y_i)$, $\dot{z}_i = 0.2 +
z_i(x_i - 9.0) + c\sum_{j=1}^{N}P_{ij}(z_j-z_i)$; (iii)
\emph{Nonidentical R\"ossler dynamics} \cite{NRossler:1997}
(N-R\"ossler): $\dot{x}_i=-\omega_i y_i-z_i +
c\sum_{j=1}^{N}P_{ij}(x_j- x_i) + \eta_i$, $\dot{y}_i =
\omega_ix_i + 0.2y_i + c\sum_{j=1}^{N} P_{ij}(y_j - y_i)$,
$\dot{z}_i = 0.2 + z_i(x_i - 9.0) +
c\sum_{j=1}^{N}P_{ij}(z_j-z_i)$, where $\omega_i$ governs the
natural frequency of an individual oscillator $i$ and  is randomly
chosen from a range $[a_1,a_2]$; (iv) \emph{Kuramoto phase
oscillators} \cite{Kuramoto:book}: $\dot{\theta_i} = \omega_i +
c\sum_{j=1}^{N} P_{ij}\sin (\theta_j - \theta_i) + \eta_i$, where
$\theta_i$ and $\omega_i$ are the phase and natural frequency of
node $i$.

Numerical simulations are carried out to predict the {\em entire}
network structure based solely on time series, utilizing the
one-to-one correspondence between the dynamical correlation and
Laplacian matrix of topology. From Eq.~(\ref{eq:undirect}), we
have $\mathbf{\hat{L}} = [\sigma^2/(2c)]\mathbf{\hat{C}}^\dag$,
where $\mathbf{\hat{L}}$ contains full information about the
network topology, and $\mathbf{\hat{C}}^\dag$ is the pseudo
inverse. The matrix $\mathbf{\hat{C}}$ can be obtained from time
series as $C_{ij} = \langle [\mathrm{x}_i(t) -
\mathrm{\bar{x}}(t)]\cdot [\mathrm{x}_j(t) - \mathrm{\bar{x}}(t)]
\rangle$, where $\mathrm{\bar{x}}(t)=
(1/N)\sum_{i=1}^{N}\mathrm{x}_i(t)$. For Kuramoto oscillators,
$\mathrm{x}_i(t)$ denotes the phase variable $\theta (t)$ and for
the R\"ossler dynamics, $\mathrm{x}_i(t)$ is the $x$ component of
the $i$th oscillator \cite{note:xt}. After $\mathbf{\hat{C}}$ is
constructed, we are able to obtain $\mathbf{\hat{L}}$ through the
pseudo inverse.

\begin{table}
\caption{Success rates of existent links (SREL) and of
non-existent links (SRNL) \cite{rate} with our method for (i)
Consensus, (ii) I-R\"ossler, (iii) N-R\"ossler, and (iv) Kuramoto
dynamics on random \cite{ER:1959}, small-world \cite{WS:1998},
scale-free model networks \cite{BA:1999}, and six real-world
networks: network of political book purchases (Book) \cite{PBP},
friendship network of karate club (Karate) \cite{FKC}, network of
American football games among colleges (Football) \cite{ACF},
electric circuit networks (Elec. Cir.) \cite{EC}, dolphin social
network (Dolphins) \cite{DS}, and the neural network of C. Elegans
(C. Elegans) \cite{WS:1998}. The noise strength is $\sigma^2=2$.
For the non-identical R\"ossler system, $\omega = [0.8,1.2]$ and
for the Kuramoto dynamics, $\omega = [0,0.2]$. Other parameters of
model networks are the same as Fig.~\ref{fig:distribution}. }
\label{tab:data1}
\begin{center}
\begin{tabular}{|c|c|c|c|c|}
  \hline
  SREL/SRNL  & consensus & I-R\"ossler & N-R\"ossler & Kuramoto \\
  \hline
  Random & 1.00/1.00 & 1.00/1.00 & 0.995/1.00 & 0.977/0.999 \\
  \hline
  Small-world & 0.993/1.00 & 0.988/1.00 & 0.979/1.00 & 0.982/1.00 \\
  \hline
  Scale-free & 0.995/1.00 & 0.990/1.00 & 0.980/1.00 & 0.978/1.00 \\
  \hline
  Book & 0.971/1.00 & 0.977/1.00 & 0.964/1.00 & 0.967/1.00 \\
  \hline
  Karate & 0.962/1.00 & 0.962/1.00 & 0.936/1.00 & 0.949/1.00 \\
  \hline
  Football & 0.938/1.00 & 0.932/1.00 & 0.928/1.00 & 0.927/1.00 \\
  \hline
  Elec. Cir. & 0.976/1.00 & 0.973/1.00 & 0.971/1.00 & 0.965/1.00 \\
  \hline
  Dolphins & 0.984/1.00 & 0.981/1.00 & 0.984/1.00 & 0.973/1.00 \\
  \hline
  C. Elegans & 1.00/0.997 & 1.00/0.996 & 1.00/0.997 & 0.993/0.997 \\
  \hline
\end{tabular}
\end{center}
\end{table}

\begin{table}
\caption{SREL with our method for consensus and N-R\"ossler
dynamics on random, small-world, scale-free networks with
different average degree $\langle k\rangle$. SRNL for all cases
are $1.000$ (not shown). Parameters are the same as
Table~\ref{tab:data1}.}\label{tab:data2}
\begin{center}
\begin{tabular}{|c|c c c| c c c|}
 \hline
  SREL  & & consensus  &  &  & N-R\"ossler & \\
 \hline
 $\langle k\rangle$ & 8 & 10 & 12 & 8 & 10 & 12 \\
 \hline
 Random & 0.986 & 0.993 & 0.996 & 0.975 & 0.984 & 0.989 \\
 \hline
 Small-world & 0.952 & 0.977 & 0.993 & 0.935 & 0.966 & 0.977 \\
 \hline
 Scale-free & 0.986 & 0.995 & 0.997 & 0.964 & 0.980 & 0.987 \\
\hline
\end{tabular}
\end{center}
\end{table}

Figure \ref{fig:distribution} shows the distribution of elements
of $[\sigma^2/(2c)]\mathbf{\hat{C}}^{\dag}$. We observe a bimodal
distribution with one peak centered at $-1$ corresponding to
existent links and the other peak centered at zero corresponding
to zero elements in $\mathbf{\hat{L}}$. There are also some
positive values in the distribution that disperse on the right
side of the peak about zero, which are due to the diagonal
components in $\mathbf{\hat{L}}$. We focus on non-diagonal
elements in $\mathbf{\hat{L}}$. If $\mathbf{\hat{L}}$ were
reconstructed perfectly from
$[\sigma^2/(2c)]\mathbf{\hat{C}}^{\dag}$, the two peaks would be
very sharp. A threshold can be set to distinguish existent from
non-existent links by using Eq.~(\ref{eq:local}). In particular,
from Eq.~(\ref{eq:local}), we have $S \equiv \sum_{i=1}^{N}
1/C_{ii} = 2cl^2/[\sigma^2 (N+l)]$, where $l=\sum_{i=1}^{N}k_i
=N\langle k\rangle$ is twice the total number of links. We can
calculate $l$ through $l=(S\sigma^2 + \sqrt{S^2 \sigma^4 +
8cNS\sigma^2 })/4c$ and keep its integral part. We then rank all
elements of the matrix $\mathbf{\hat{C}}^{\dag}$ (or matrix
$[\sigma^2/(2c)]\mathbf{\hat{C}}^{\dag}$) in an ascending order.
For convenience, we denote the ascending-ordered matrix elements
by $C_m^{\dag}$, for $m = 1, \ldots, N^2$. The threshold
$C_M^{\dag}$ (or $[\sigma^2/(2c)]C_M^{\dag}$) is chosen such that
$\sum^M_{m=1} \Phi(C_m^{\dag}) = l$, where $\Phi(C_m^{\dag})$ is
the unnormalized distribution of $C_m^{\dag}$. This means the rank
of $C_M^{\dag}$ is $l$ in the queue of ascending-ordered matrix
elements $C_m^{\dag}$. Then the connection matrix can be obtained
by setting all elements in $\mathbf{\hat{C}}^\dag$ with values
above the threshold $C_M^{\dag}$ to be zero and others to be $-1$,
the latter corresponding to existent links. As shown in
Fig.~\ref{fig:distribution}, for different model and real-world
networks, thresholds so determined are able to successfully
separate the two peaks in the distributions of elements of
$\mathbf{\hat{C}}^{\dag}$, which in turn leads to predictions of
links with high success rates for various node dynamics, as
displayed in Table~\ref{tab:data1}. Alternatively, the threshold
can be empirically determined by the largest gap between the two
peaks, and we have obtained essentially the same success rates.
Table~\ref{tab:data2} exemplifies the success rates of our method
for different values of the average degree $\langle k\rangle$ for
different types of networks. We see that the success rate
increases with $\langle k\rangle$.

\begin{figure}
\begin{center}
\epsfig{figure=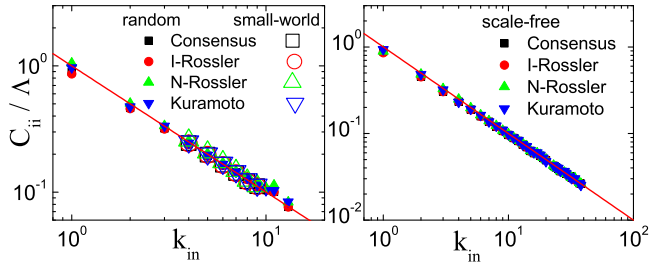,width=\linewidth} \vspace{-.5cm}
\caption{(Color online) $C_{ii}$ as a function of node in-degree
$k_{in}$ for different node dynamics for directed networks where
each link is assigned a random direction, and $\Lambda =
{\sigma}^2(1+1/\langle k\rangle)/2c$. Other network parameters are
the same as in Fig.~\ref{fig:distribution}. The lines are
predictions from Eq.~(\ref{eq:local}).} \label{fig:cii}
\end{center}
\end{figure}

For directed networks, there is no unique solution for
$\mathbf{\hat{L}}$ from $\mathbf{\hat{C}}$, because the asymmetric
$\mathbf{\hat{L}}$ has a twofold degree of freedom as that of
symmetric $\mathbf{\hat{C}}$. Thus, the global structure of
directed networks cannot be inferred solely depending on the
correlation. However, Eq.~(\ref{eq:local}) is satisfied by
replacing node degree with in-degree, so that we still can infer
the local structure, the in-degree of each node, through
$k_{in}^i\sim C_{ii}^{-1}$. As shown in Fig.~\ref{fig:cii}, theory
agrees well with numerical results.

In conclusion, we have discovered a general relation between the
dynamical correlation among oscillators and the underlying
topology in the presence of noise. The correlation matrix is
inversely proportional to the Laplacian matrix that contains full
information about the network structure. Reconstruction of the
full network topology based on time series then becomes possible,
particularly for undirected networks. We have provided strong
numerical support by using four types of nodal dynamics together
with several model and real-world network structures. We find that
the full network topology can be predicted with high success rate
and efficiency for all considered cases. Besides high success
rates, advantages making our method attractive and powerful
include generality for a variety of nodal dynamics and network
structures, validity in the existence of weak coherence,
applicability in the absence of knowledge about nodal dynamics,
and no need to control nodal dynamics as in some existing method.
We hope that our method can be widely applied for inferring
network structures and inspire further research towards the
understanding of noise effects on networked dynamical systems.

We thank anonymous referees for valuable suggestions on
theoretically determining thresholds. WXW and YCL are supported by
AFOSR under Grant No. FA9550-07-1-0045.


\begin{references}
\bibitem{note:green}
We can decompose $\mathbf{\hat{L}}$ into two parts:
$\mathbf{\hat{L}} = \mathbf{\hat{K}} - \mathbf{\hat{P}}$, where
$\mathbf{\hat{P}}$ is the adjacency matrix with $P_{ij} = 1$ if
node $j$ connects to $i$ (otherwise $0$), and $\mathbf{\hat{K}} =
\mathrm{diag}(k_1,\cdots , k_N)$, where $k_i$ is the degree of
node $i$. The matrix $\mathbf{\hat{C}}$ can thus be expressed in a
series: $\mathbf{\hat{C}} \sim (\mathbf{\hat{K}} -
\mathbf{\hat{P}})^{-1} \nonumber = \mathbf{\hat{K}}^{-1} +
\mathbf{\hat{K}}^{-1}\mathbf{\hat{P}}\mathbf{\hat{K}}^{-1} +
\mathbf{\hat{K}}^{-1}\mathbf{\hat{P}}\mathbf{\hat{K}}^{-1}\mathbf{\hat{P}}\mathbf{\hat{K}}^{-1}+
\cdots$. 
For the second term
$\mathbf{\hat{K}}^{-1}\mathbf{\hat{P}}\mathbf{\hat{K}}^{-1}$, if
nodes $j$ connects to $i$, its element $(i,j)$ is $(k_i k_j)^{-1}$
and otherwise 0. For the third term
$\mathbf{\hat{K}}^{-1}\mathbf{\hat{P}}\mathbf{\hat{K}}^{-1}\mathbf{\hat{P}}\mathbf{\hat{K}}^{-1}$,
if there are multiple two-step paths connecting $j$ to $i$ through
node $m_1$ or $m_2$, $\cdots$, $m_r$, its element $(i,j)$ values
$k^{-1}_i\cdot(\sum_{q=m_1}^{m_r}k^{-1}_q)\cdot k^{-1}_j$.
\bibitem{note:xt}
$\mathrm{x}_i(t)$ can also be the $y$ or $z$ component with noise
in R\"ossler dynamics. We have examined the success rate by using
the time series of $y$ or $z$ for various networks. High success
rates are obtained as well, similar to the results by using the
$x$-component.
\bibitem{rate} SREL (SRNL) is defined as the ratio of
the number of successfully predicted existent (non-existent) links
to the total number of existent (non-existent) links.
\end{references}
\end{document}